\documentclass[twocolumn,showpacs,preprintnumbers,amssymb]{revtex4}

\usepackage{graphicx}% Include figure files
\usepackage{bm}
\begin{document}

\centerline{}
%\vskip 1cm
\title{Quasinormal modes of Schwarzschild black holes in four and higher 
dimensions}

\author{Vitor Cardoso}
\email{vcardoso@fisica.ist.utl.pt}
\author{Jos\'e P. S. Lemos}
\email{lemos@kelvin.ist.utl.pt} 
\author{Shijun Yoshida}
\email{yoshida@fisica.ist.utl.pt}
\affiliation{
Centro Multidisciplinar de Astrof\'{\i}sica - CENTRA, 
Departamento de F\'{\i}sica, Instituto Superior T\'ecnico,
Av. Rovisco Pais 1, 1049-001 Lisboa, Portugal,}

\date{\today}

\begin{abstract}

We make a thorough investigation of the asymptotic quasinormal modes
of the four and five-dimensional Schwarzschild black hole for scalar,
electromagnetic and gravitational perturbations.  Our numerical
results give full support to all the analytical predictions by Motl
and Neitzke, for the leading term.  We also compute the first order
corrections analytically, by extending to higher dimensions, 
previous work of Musiri and
Siopsis, and find excellent agreement with the
numerical results.  For generic spacetime dimension number $D$ the
first-order corrections go as $\frac{1}{n^{(D-3)/(D-2)}}$. This means
that there is a more rapid convergence to the asymptotic value
for the five dimensional case than for the four dimensional case, 
as we also show numerically.

\end{abstract}

%\pacs{}

\maketitle
\newpage
%%%%%%%%%%%%%%%%%%%%%%%%%%%%%%%%%%%%%%%%%%%%%%%%%%%%%
\section{Introduction}
%%%%%%%%%%%%%%%%%%%%%%%%%%%%%%%%%%%%%%%%%%%%%%%%%%%%%
The study of quasinormal modes of black holes began more than thirty
years ago, when Vishveshwara \cite{vish} noticed that the signal from
a perturbed black hole is, for most of the time, an exponentially
decaying ringing signal. It turns out that the ringing frequency and
damping timescale are characteristic of the black hole, depending only
on its parameters (like the mass and angular momentum).  We call these
characteristic oscillations the quasinormal modes (QNMs) and the
associated frequencies are termed quasinormal frequencies (QN
frequencies), because they are really not stationary perturbations.  
Not surprisingly, QNMs play an important role in the
dynamics of black holes, and consequently in gravitational wave
physics. In fact, it is possible \cite{echeverria,nakanoringing} to
extract the parameters of the black hole simply by observing these
QN frequencies, using for example gravitational wave
detectors.  The discovery that QNMs dominate the answer of a black
hole to almost any exterior perturbation was followed by a great
effort to find, numerically and analytically, the QN frequencies.
For excellent reviews on the status of QNMs, prior to 2000,
we refer the reader to Kokkotas and Schmidt
\cite{kokkotas} and Nollert \cite{nollert}.
It is important to note that on the astrophysical aspect, the most
important QN frequencies are the lowest ones, i.e., frequencies 
with smaller imaginary part, and the most important spacetimes are the 
asymptotically flat
and perhaps now the asymptotically de Sitter.  
However, three years ago Horowitz and Hubeny \cite{horowitz}
pointed out that QNMs of black holes in anti-de Sitter space have a
different importance.  According to the AdS/CFT correspondence
\cite{maldacena}, a black hole in anti-de Sitter space may be viewed
as a thermal state in the dual theory. Perturbing this black hole
corresponds to perturbing the thermal state, and therefore the typical
timescale of approach to thermal equilibrium (which is hard to compute
directly in the dual theory) should as well be governed by the  lowest
QN frequencies.  That this is indeed the case was proved by Birmigham,
Sachs and Solodukhin \cite{birmingham2} for the BTZ black hole,
taking advantage that this is one of the few spacetimes where one can
compute exactly its QN frequencies, as showed by Cardoso and Lemos
\cite{cardosoqnmbtz}.  A similar study was made by Kurita and Sakagami
\cite{kurita} for the D-3 brane.
This interpretation for the imaginary part of the QN frequencies 
in terms of timescales of approach
to thermal equilibrium in the dual conformal field theory has
motivated a generalized search for the quasinormal modes of different
black holes in anti-de Sitter spacetime, over the last three years 
\cite{horowitz,cardosoqnmbtz,qnmads}.

Recently, the motivation for studying QN modes of black holes has grown
enormously with the conjectures \cite{hod,dreyer,oppenheim,ling}
relating the highly damped QNMs (i.e., QN frequencies with large
imaginary parts) to black hole area quantization and to the Barbero
Immirzi parameter appearing in Loop Quantum Gravity. The seeds of that idea
were planted some
time ago by Bekenstein \cite{bek0}. A semi-classical reasoning of
the conjecture \cite{bek1,bek2} that the black hole area spectrum
is quantized leads to
\begin{equation}
A_{\,n}=\gamma\, l_{P}^2\, n\,,\quad n=1,2,...\quad.
\label{areaspectrum}
\end{equation}
Here $l_{P}$ is the Planck length and $\gamma$ is an undetermined constant.
However, statistical physics arguments impose a constraint on $\gamma$ 
\cite{bek1,bek3}:
\begin{equation}
\gamma=4 \log k\,,
\label{areaspectrum1}
\end{equation}
where $k$ is an integer.
The integer $k$ was left undetermined (although there were some suggestions
for it \cite{bek2}), until Hod \cite{hod}, supported
by some of Bekenstein's ideas, put forward the proposal to determine
$k$ via a version of Bohr's correspondence principle, in which one admits
that the real part of QN frequencies with a large imaginary
part plays a fundamental role. 
It was seen numerically by Nollert \cite{nollert,nollert2}
that QN frequencies with a large imaginary part behave,
in the Schwarzschild geometry as
\begin{equation}
\omega M= 0.0437123+\frac{i}{8}\,(2n+1)\,,
\label{asymptQNnollert}
\end{equation}
where $M$ is the black hole mass, and $n$ the mode number.
Hod first realized that $ 0.0437123 \sim \frac{\ln3}{8\pi}$, and 
went on to say that, if one supposes that the emission of a quantum
with frequency $\frac{\ln3}{8\pi}$ corresponds to the least possible energy 
a black hole can emit, then the change in surface area will be
(using $A=16\pi M^2$)
\begin{equation}
\Delta A= 32\pi MdM=32\pi M{\hbar} \omega= 4{\hbar}\ln 3 \,.
\label{surfacearea}
\end{equation}
Comparing with (\ref{areaspectrum}) we then get $k=3$ and therefore the area
spectrum is fixed to
\begin{equation}
A_{\rm n}=4\ln 3 l_{P}^2 n\,;\,\,n=1,2,...
\label{areaspectrumfinal}
\end{equation}
It was certainly a daring proposal to map $0.0437123$ to $\ln
3$, and even more to use the QN frequencies to quantize the black hole
area, by appealing to ``Bohr's correspondence principle''. The
risk paid off: recently, Dreyer \cite{dreyer} put forward the
hypothesis that if one uses such a correspondence it is possible to
fix a formerly free parameter, the Barbero-Immirzi parameter, appearing
in Loop Quantum Gravity. Dreyer's work implied that 
the entropy has contributions from $J=1$ edges mainly.
One interpretation (proposed by Dreyer), is that one changes
the gauge group from $SU(2)$, which is the most natural one, to
$SO(3)$.
There are other explanations, though.  For example, Corichi
\cite{corichi} presents an argument based on simple conservation
principles to suggest that one can keep $SU(2)$ and still have
consistency with QNM, while Ling and Zhang \cite{ling} propose to
consider the supersymmetric extension of the theory.

All of these proposals and conjectures could be useless
if one could not prove that the real part of the QN frequencies
do approach $\frac{\ln3}{8\pi}$, i.e., that the number $0.0437123$ in
Nollert's paper was exactly $\frac{\ln3}{8\pi}$. This
was accomplished by Motl \cite{motl1} some months ago, using an
ingenious technique, working with the continued fraction
representation of the wave equation.  Subsequently, Motl and Neitzke
\cite{motl2} used a more flexible and powerful approach, called 
the monodromy method, and were not only able to rederive the four
dimensional Schwarzschild value $\frac{\ln3}{8\pi}$, but also to
compute the asymptotic value for generic $D$-dimensional Schwarzschild
black holes thereby confirming a previous conjecture by Kunstatter
\cite{kunstatter}.
They have also predicted the form of the asymptotic QN frequencies
for the Reissner-Nordstr\"om geometry, a prediction which was verified
by Berti and Kokkotas \cite{bertirn}.

It is interesting to speculate why, in thirty years investigating QNMs
and QN frequencies, no one has ever been able to deduce analytically
the asymptotic value of the QN frequencies, nor even for the
Schwarzschild geometry, which is the simplest case.  In our opinion,
there are three important facts that may help explain this: first,
the concern was mostly with the lowest QN frequencies, the ones
with smaller imaginary part, since they are the most important in the
astrophysical context where QNMs were inserted until three years
ago. In fact, most of the techniques to find QN frequencies, and there
were many, have been devised to compute the lowest QN frequencies (see
\cite{kokkotas,nollert} for a review).  Secondly, there was no 
suggestion for the asymptotic $\ln{3}$ value. 
To know this value serves as
a strong stimulation. The third reason is obvious: this was a very difficult
technical problem.

It is thus satisfying that Motl and Neitzke have proved exactly
that the limit is $\ln{3}$. In addition, Van den Brink \cite{brinkasympt} 
used a
somewhat different approach to rederive the $\frac{\ln3}{8\pi}$
value, and at the same time computed the first order correction to
gravitational QN frequencies, confirming analytically Nollert's
\cite{nollert2} numerical results.  However,
Motl and Neitzke's monodromy method is more flexible and powerful: it
has been easily adapted to other spacetimes.  For example, Castello-Branco 
and Abdalla \cite{karlucio} have generalized the results to the
Schwarzschild-de Sitter geometry, while Schiappa, Nat\'ario and
Cardoso \cite{ricardo} have used it to compute the asymptotic values
in the $D$-dimensional Reissner-Nordstr\"om geometry.
Moreover the first order corrections are also easily obtained, as has
been showed by Musiri and Siopsis \cite{musiri2} for the four
dimensional Schwarzschild geometry. Here we shall also generalize
these first order corrections so that they can encompass a general
$D$-dimensional Schwarzschild black hole.
A body of work has been growing on the subject of computing highly
damped QNMs, both in asymptotically flat and in de Sitter or anti-de
Sitter spacetimes \cite{qnmkerr,asymptall}.

As an aside, we note that one of the most intriguing
features of Motl and Neitzke's technique is that the region beyond the
event horizon, which never enters in the definition of quasinormal
modes, plays an extremely important role.  In particular, the
singularity at $r=0$ is decisive for the computation.  At the
singularity the effective potential for wave propagation blows
up. Somehow, the equation, or the singularity, knows what we are
seeking! It is also worth of note the following: for high frequencies, or
at least frequencies with a large imaginary part, the important region
is therefore $r=0$ where the potential blows up, whereas for low
frequencies, of interest for late-time tails \cite{tails}, it is the
other limit, $r \rightarrow \infty$ which is important.

The purpose of this paper is two-fold: first we want to enlarge and settle
the results for the four dimensional Schwarzschild black hole.  We shall
first numerically confirm Nollert's results for the gravitational
case, and Berti and Kokkotas'\cite{bertirn} results for the scalar
case, and see that the match between these and the analytical
prediction is quite good. Then we shall fill-in a gap left behind in
these two studies: the electromagnetic asymptotic QN frequencies, for
which there are no numerical results, but a lot of analytical
predictions.  This will put the monodromy method on a firmer
ground.  One might say that in the present context (of black hole
quantization and relation with Loop Quantum Gravity) electromagnetic
perturbations are not relevant, only the gravitational ones are. This
is not true though, because when considering the Reissner-Nordstr\"om case
gravitational and electromagnetic perturbations are coupled
\cite{chandra} and it is therefore of great theoretical interest to
study each one separately, in the case they do decouple, i.e., the
Schwarzschild geometry.
We find that all the analytical predictions are correct: scalar and
gravitational QN frequencies asymptote to 
$\frac{\omega}{T_H}=\ln{3}+i(2n+1)\pi$ 
with $\frac{1}{\sqrt n}$ corrections, where 
$T_H=\frac{1}{8\pi M}$ is the Hawking temperature.  The
corrections are well described by the existing analytical formulas.
On the other hand, electromagnetic QN frequencies asymptote to
$\frac{\omega}{T_H}=0+i2n\pi$ with no $\frac{1}{\sqrt n}$ corrections, as
predicted by Musiri and Siopsis \cite{musiri2}.  Indeed we find that
the corrections seem to appear only at the $\frac{1}{n^{3/2}}$ level.

The second aim of this paper is to establish numerically the validity of the
monodromy method by extending the numerical calculation to higher
dimensional Schwarzschild black holes, in particular to the 
five dimensional one.  We obtain numerically in the five dimensional case
the asymptotic
value of the QN frequencies, which turns out to be $\ln 3+i(2n+1)\pi$ (in
units of the black hole temperature), for scalar, gravitational tensor
and gravitational vector perturbations.  Equally important, is that
the first order corrections appear as $\frac{1}{n^{2/3}}$ for these
cases.  We shall generalize Musiri and Siopsis' method to higher
dimensions and find that (i) the agreement with the numerical data is
very good (ii) In generic $D$ dimensions the corrections appear at the
$\frac{1}{n^{(D-3)/(D-2)}}$ level.  Electromagnetic perturbations in
five dimensions seem to be special: according to the analytical
prediction by Motl and Neitzke, they should not have any asymptotic QN
frequencies. Our numerics seem not to go higher enough in mode number
to prove or disprove this. They indicate that the real part approaches
zero, as the mode number increases, but we have not been able to
follow the QN frequencies for very high overtone number.  We also give
complete tables for the lowest lying QN frequencies of the five
dimensional black holes, and confirm previous results
\cite{vitoroscarjose,konoplyawkb,roman2,ida} for the fundamental QN
frequencies obtained using WKB-like methods.  We find no purely
imaginary QN frequencies for the gravitational QNMs, which may be
related to the fact that the potentials are no longer superpartners.

%%%%%%%%%%%%%%%%%%%%%%%%%%%%%%%%%%%%%%%%%%%%%%%%%%%%%%%%%%%%%%%%%%%%%%%%%%%
\section{QN frequencies of the four-dimensional Schwarzschild black hole }
%%%%%%%%%%%%%%%%%%%%%%%%%%%%%%%%%%%%%%%%%%%%%%%%%%%%%%%%%%%%%%%%%%%%%%%%%%%

The first computation of QN frequencies of four-dimensional
Schwarzschild black holes was carried out by Chandrasekhar and
Detweiler \cite{chandradet}.  They were only able to compute the
lowest lying modes, as it was and still is quite hard to find
numerically QN frequencies with a very large imaginary part.  After
this pioneering work, a number of analytical, semi-analytical and
numerical techniques were devised with the purpose of computing higher
order QNMs and to establish the low-lying ones using different methods
\cite{kokkotas,nollert}. The most
successful numerical method to study QNMs of black holes was proposed
by Leaver \cite{Le85}. In a few words his method amounts to reducing the
problem to a continued fraction, rather easy to implement
numerically. Using this method Leaver was able to determine the first
60 modes of the Schwarzschild black hole, which at that time was about
an order of magnitude better than anyone had ever gone before.  The
values of the QN frequencies he determined, and we note he only
computed gravitational ones, suggested that there is an infinite
number of QN modes, that the real part of the QN frequencies approach
a finite value, while the imaginary part grows without bound.  
That the real part of the QN frequencies approach
a finite value was however opened to debate \cite{guinn}: it seemed one still 
had to
go higher in mode number in order to ascertain the true asymptotic
behaviour.  Nollert \cite{nollert2} showed a way out
of these problems: he was able to improve Leaver's method in a very
simple fashion so as to go very much higher in mode number: he was
able to compute more than 2000 modes of the gravitational QNMs.  His
results were clear: the real part of the gravitational QN frequencies
approaches a constant value $\sim 0.0437123/M $, and this value is
independent of $l$. The imaginary part grows without bound as
$(2n+1)/8$ where $n$ is any large integer.  These asymptotic
behaviours have both leading corrections, which were also determined
by Nollert. He found that to leading order the asymptotic behaviour
of the gravitational QN frequencies of a Schwarzschild black hole is
given by Eq. (\ref{asymptQNnollert}) plus a leading correction, 
namely,
\begin{equation}
\omega M=0.0437123+\frac{i}{8}\,(2n+1)+ \frac{a}{\sqrt{n}}\,,
\label{nono}
\end{equation}
where the correction term $a$ depends on $l$.  About ten years
later, Motl \cite{motl1} proved analytically that this is indeed the
correct asymptotic behaviour. To be precise he proved that
\begin{equation}
\frac{\omega}{T_H}=\ln{3}+i\,(2n+1)\pi+\,{\rm corrections}\,.
\label{nonolubos}
\end{equation}
Since the Hawking temperature $T_H$ is 
$T_H=\frac{1}{8\pi M}$, Nollert's
result follows.
He also proved that this same result also holds for scalar QN
frequencies, whereas electromagnetic QN frequencies asymptote to a
zero real part.  Motl and Neitzke \cite{motl2} using a completely
different approach, have rederived this result whereas Van den Brink
\cite{brinkasympt} was able to find the leading asymptotic correction
term $a$ in (\ref{nono}) for the gravitational case.  Very
recently, Musiri and Siopsis \cite{musiri2}, leaning on Motl and
Neitzke's technique, have found the leading correction term for any
field, scalar, electromagnetic and gravitational.
This correction term is presented in section \ref{apendice}, where
we also generalize to higher dimensional black holes. 
In particular, for electromagnetic perturbations they find that the 
first order
corrections are zero.

Here we use Nollert's method to compute scalar, electromagnetic and
gravitational QN frequencies of the four-dimensional Schwarzschild
black hole.  The details on the implementation of this method in four
dimensions are well known and we shall not dwell on it here any more.
We refer the reader to the original references \cite{nollert2,Le85}.
\begin{figure}
\centerline{\includegraphics[width=7 cm,height=7 cm]
{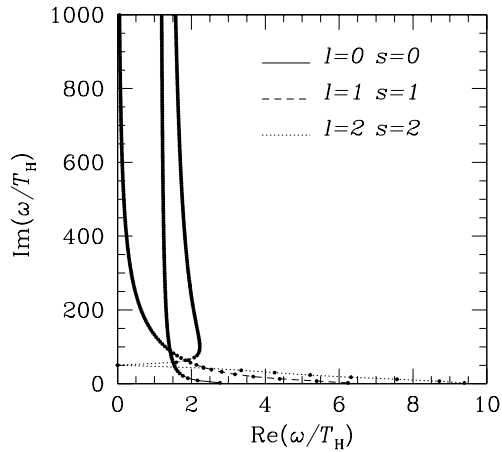}}
\caption{Scalar ($s=0$), electromagnetic ($s=1$) and gravitational ($s=2$)
QN frequencies of a four-dimensional Schwarzschild black hole. 
Note that the dots indicate the QN frequencies there, and the lines
connecting the dots only help to figure out
to which multipole ($l$, $s$) they belong to. 
We show
the lowest 500 modes for the lowest radiatable multipoles of each
field, i.e., $l=0$ for the scalar field, $l=1$ for the electromagnetic
and $l=2$ for gravitational.  However, the asymptotic behaviour is
$l$-independent. It is quite clear from this plot that highly damped
electromagnetic QN frequencies have a completely different behaviour
from that of scalar or gravitational highly damped QN frequencies. In
fact, the electromagnetic ones asymptote to $\frac{\omega}{T_H} \rightarrow
i(2n)\pi\,\,,n\rightarrow \infty$, as predicted by Motl
\cite{motl1} and Motl and Neitzke \cite{motl2}. The scalar and
gravitational ones asymptote to $\frac{\omega}{T_H}\rightarrow
\ln{3}+i(2n+1)\pi\,\,,n\rightarrow \infty$.}
\label{fig:qnall4d}
\end{figure}

\begin{figure}
\centerline{\includegraphics[width=7 cm,height=7 cm]
{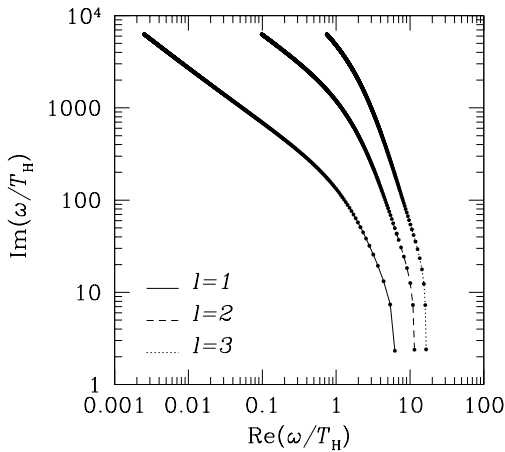}}
\caption{Electromagnetic QN frequencies of  a 
four-dimensional Schwarzschild black hole.}
\label{fig:qnelec4d}
\end{figure}

Our results are summarized in Figures
\ref{fig:qnall4d}-\ref{fig:qnelec4d}, and Tables \ref{tab:corrd4j0}-
\ref{tab:corrd4j2}.  
In Fig. \ref{fig:qnall4d} we
show the behaviour of the 500 lowest QN frequencies for the lowest
radiatable multipole of each field 
(we actually have computed the first 5000 QN frequencies, but for 
a better visualization we do not show them all): $l=0$ for scalar, $l=1$ for
electromagnetic and $l=2$ gravitational.  
Note that in Fig. \ref{fig:qnall4d}, as well as in all other plots in this 
work, 
the dots indicate the QN frequencies there, and the lines
connecting the dots only help to figure out
to which multipole ($l$, $s$) they belong to. 
In what concerns the asymptotic behaviour, our numerics show it is not
dependent on $l$, and that formula (\ref{nono}) holds. The results for
the scalar and gravitational QN frequencies are not new: the
gravitational ones have been obtained by Nollert \cite{nollert2}, as
we previously remarked, and the scalar ones have been recently arrived
at by Berti and Kokkotas \cite{bertirn}.  Here we have verified both.
Both scalar and gravitational QN frequencies have a real part
asymptoting to $\ln{3}\times T_H$.  Electromagnetic QN frequencies,
on the other hand, have a real part asymptoting to $0$. This is
clearly seen in Fig. \ref{fig:qnall4d}.  For gravitational
perturbations Nollert found, and we confirm, that the correction term
$a$ in (\ref{nono}) is
\begin{eqnarray}
a= 0.4850\,,\,\,l=2
\nonumber \\
\!\!\!a= 1.067 \,,\,\,l=3
\label{corrtermsnum}
\end{eqnarray}
We have also computed the correction terms for the scalar case, and we have 
obtained
results in agreement with Berti and Kokkotas' ones \cite{bertirn} .
Our results are summarized in Tables \ref{tab:corrd4j0}-\ref{tab:corrd4j2}, 
where
we also show the analytical prediction \cite{musiri2} (see section 
\ref{apendice}).
To fix the conventions adopted throughout the rest of the paper,
we write
\begin{equation}
\frac{\omega}{T_H}=\ln{3}+i(2n+1)\pi+\frac{{\rm Corr}}{\sqrt{n}}\,.
\label{conv}
\end{equation}
Thus, the correction term $a$ in (\ref{nono}) is related to {\rm Corr}
by $a={\rm Corr}\times T_H$. 

\vskip 1mm
\begin{table}
\caption{\label{tab:corrd4j0} The correction coefficients for the
four-dimensional Schwarzschild black hole, both numerical, here
labeled as ``${\rm Corr_{4}^{N}}$'' and analytical, labeled as ``${\rm
Corr_{4}^{A}}$''. These results refer to scalar
perturbations.
The analytical results are extracted from \cite{musiri2}
(see also formula (\ref{corrd4} below).  Notice the very good agreement
between the numerically extracted results and the
analytical prediction.These values were obtained using the first
5000 modes.}
\begin{ruledtabular}
\begin{tabular}{l|ll}  \hline
\multicolumn{1}{c}{} &
\multicolumn{2}{c}{ $s=0$}\\ \hline
$l$ &${\rm Corr_4^N}$:&${\rm Corr_4^A}$\\ \hline
0   &1.20-1.20i  &1.2190-1.2190i  \\ \hline 
1   &8.52- 8.52i  &8.5332-8.5332i   \\ \hline 
2   &23.15-23.15i  &23.1614-23.1614i   \\ \hline 
3   &44.99-44.99i  &45.1039-45.1039i \\ \hline 
4   &74.34-74.34i &74.3604-74.3605i  \\ \hline 
\end{tabular}
\end{ruledtabular}
\end{table}
\vskip 1mm
\vskip 1mm
\begin{table}
\caption{\label{tab:corrd4j2} The correction coefficients for the
four-dimensional Schwarzschild black hole, both numerical, here
labeled as ``${\rm Corr_{4}^{N}}$'' and analytical, labeled as ``${\rm
Corr_{4}^{A}}$''. These results refer to gravitational
perturbations.
The analytical results are extracted from \cite{musiri2}
(see also formula (\ref{corrd4} below). Notice the very good agreement
between the numerically extracted results and the
analytical prediction. These values were obtained using the first
5000 modes.}
\begin{ruledtabular}
\begin{tabular}{l|ll}  \hline
\multicolumn{1}{c}{} &
\multicolumn{2}{c}{ $s=2$}\\ \hline
$l$ &${\rm Corr_4^N}$:&${\rm Corr_4^A}$\\ \hline
2   & 6.08-6.08i   & 6.0951-6.0951i  \\ \hline 
3   & 13.39-13.39i & 13.4092-13.4092i \\ \hline 
4   & 23.14-23.14i & 23.1614-23.1614i \\ \hline 
5   & 35.33-35.33i & 35.3517-35.3517i\\ \hline 
6   & 48.90-48.90i & 49.9799-49.9799i \\ \hline 
\end{tabular}
\end{ruledtabular}
\end{table}
\vskip 1mm
As for the electromagnetic correction terms, we found it was not very easy 
to determine
them. In fact, it is hard, even using Nollert's method, to go very high in mode
number for these perturbations (we have made it to $n=5000$).
The reason is tied to the fact that these frequencies asymptote to zero,
and it is hard to determine QN frequencies with a vanishingly small real part.
Nevertheless, there are some features one can be sure of.
We have fitted the electromagnetic data to the following form
$\frac{\omega}{T_H}=i(2n)\pi+\frac{b}{\sqrt{n}}\,,$
and found that this gave very poor results. Thus one saw numerically that
the first order correction term is absent, as predicted in \cite{musiri2}
(see also section \ref{apendice} where we rederive these corrections, 
generalizing them
to arbitrary dimension).
Our data seems to indicate that the leading correction term is of the form
$\frac{b}{n^{3/2}}$. This is more clearly seen in Fig. \ref{fig:qnelec4d}
where we plot the first electromagnetic QN frequencies on a $\ln$ plot.
For frequencies with a large imaginary part, the slope is about $-2/3$ as
it should if the asymptotic behaviour is of the form
\begin{equation}
\frac{\omega}{T_H}=i2n\pi+\frac{b}{n^{3/2}}\,.
\label{true}
\end{equation}
Our numerics indicate that $b$ has an $l$-dependence going like
\begin{equation}
b \sim c_1 l(l+1)+c_2 \left[l(l+1)\right]^2+c_3 \left[l(l+1)\right]^3\,,
\label{bcoef}
\end{equation}
where $c_1$, $c_2$, and$c_3$ are constants. This is also the expected
behaviour, since a $\frac{1}{n^{3/2}}$ dependence means we have to go
to third order perturbation theory, where we get corrections of the
form (\ref{bcoef}) as one can easily convince oneself.
%%%%%%%%%%%%%%%%%%%%%%%%%%%%%%%%%%%%%%%%%%%%%%%%%%%%%%%%%%%%%%%%%%%%%%%%%%%
\section{QN frequencies of $D$-dimensional Schwarzschild black holes }
%%%%%%%%%%%%%%%%%%%%%%%%%%%%%%%%%%%%%%%%%%%%%%%%%%%%%%%%%%%%%%%%%%%%%%%%%%%

%%%%%%%%%%%%%%%%%%%%%%%%%%%%%%%%%%%%%%%%%%%%%%%%%%%%%%%%%%%%%%%%%%%%%%%%%%%%
\subsection{Equations and conventions}
%%%%%%%%%%%%%%%%%%%%%%%%%%%%%%%%%%%%%%%%%%%%%%%%%%%%%%%%%%%%%%%%%%%%%%%%%%%%%
The metric of the $D$-dimensional
Schwarzschild black hole in ($t,r,\theta_1,\theta_2,..,\theta_{D-2}$)
coordinates is \cite{tangherlini,myersperry}
\begin{equation}
ds^2= -fdt^2+
f^{-1}dr^2
+r^2d\Omega_{D-2}^2\,,
\label{metrictang} 
\end{equation}
with
\begin{equation}
f=1-\frac{m}{r^{D-3}}.
\label{fdef}
\end{equation}
The mass of the black hole is given by
$M=\frac{(D-2)\Omega_{D-2} m}{16\pi{\cal G}}$, where
$\Omega_{D-2}=\frac{2\pi^{(D-1)/2}}{\Gamma[(D-1)/2]}$ is the area of a
unit $(D-2)$ sphere, $d\Omega_{D-2}^{2}$ is the line element on
the unit sphere $S^{D-2}$, and ${\cal G}$ is Newton's constant in 
$D$-dimensions.  
In the following it will prove more useful to rescale variables so that
the form of the metric is (\ref{metrictang}) but with $f=1-\frac{1}{r^{D-3}}$,
i.e., we shall choose $m=1$, following the current fashion.
We will only consider the linearized
approximation, which means that we are considering wave fields outside
this geometry that are so weak they do not alter this
background. 
The evolution equation for a massless scalar field follows
directly from the Klein-Gordon equation (see
\cite{cardosoradadsddim} for details).  The gravitational evolution
equations have recently been derived by Kodama and Ishibashi
\cite{kodama}. There are three kinds of gravitational perturbations,
according to Kodama and Ishibashi's terminology: the scalar
gravitational, the vector gravitational and the tensor gravitational
perturbations.  The first two already have their counterparts in
$D=4$, which were first derived by Regge and Wheeler \cite{regge} and
by Zerilli \cite{zerilli}. The tensor type is a new kind appearing in
higher dimensions. However, it obeys exactly the same equation as
massless scalar fields, as can easily be seen.  Due to the complex
form of the gravitational scalar potential, we shall not deal with
it. Instead, we shall only consider the tensor and vector type of
gravitational perturbations. In any case, if the analytic results are correct,
then the gravitational scalar QN frequencies should have the same asymptotic
form as the gravitational vector and tensor QN frequencies.  
The evolution equation for the electromagnetic field in the higher
dimensional Schwarzschild geometry was arrived at for the first time
by Crispino, Higuchi and Matsas \cite{matsas}. It has recently
been rederived by Kodama and Ishibashi, in a wider context of charged
black hole perturbations.  We shall follow Kodama and Ishibashi's
terminology. According to them, there are two kinds of electromagnetic
perturbations: the vector and scalar type. If one makes the charge of
the black hole $Q=0$ in Kodama and Ishibashi's equations one recovers
the equations by Crispino, Higuchi and Matsas, although this seems to
have been overlooked in the literature. 
The evolution equation for all kinds of
fields (scalar, gravitational and electromagnetic) 
can be reduced to the second order differential equation
\begin{equation}
\frac{d^2\Psi}{dr_{*}^2}+(\omega^2-V)\Psi=0 \,,
\label{eveq}
\end{equation}
where $r$ is a function of the tortoise coordinate $r_*$, defined through 
$\frac{\partial r}{\partial
r_*}=f(r)$, and the potential can be written in compact form as
\begin{eqnarray}
V &=& f(r)
{\biggl [}
\frac{l(l+D-3)}{r^2}+\frac{(D-2)(D-4)}{4r^2}+
\nonumber \\
& &
\frac{(1-j^2)(D-2)^2}{4r^{D-1}}
{\biggr ]}
\,.
\label{potentialj}
\end{eqnarray}
The constant $j$ depends on what kind of field one is studying:
\begin{equation}
j=\left\{ \begin{array}{llll}
            0\,,   & {\rm \,scalar\, and\, gravitational\, tensor}\\
                                  &  {\rm perturbations}. \\ 
            1\,,   &{\rm \, gravitational\, vector} \\
                                     &  {\rm perturbations}. \\
            \frac{2}{D-2}\,,   &{\rm \, electromagnetic \,vector} \\
                                     &  {\rm perturbations}. \\
            2- \frac{2}{D-2}\,,   &{\rm \, electromagnetic\, scalar} \\
                                     &  {\rm perturbations}\,.
\end{array}\right.
\label{jdef}
\end{equation}
Notice that in four dimensions, $j$ reduces to the usual values
given by Motl and Neitzke \cite{motl2}. 
According to our conventions, the Hawking temperature of a
$D$-dimensional Schwarzschild black hole is
\begin{equation}
T_H=\frac{D-3}{4\pi}
\label{hawktemp}
\end{equation}

Our purpose here is to investigate the QN frequencies of higher
dimensional black holes.  Of course one cannot study every $D$. We
shall therefore focus on one particular dimension, five, and make a
complete analysis of its QN frequencies. The results should be
representative. In particular, one feature that distinguishes 
an arbitrary $D$ from the four dimensional case is that the correction
terms come with a different power, i.e., whereas in four dimensions
the correction is of the form $\frac{1}{\sqrt{n}}$ we shall find that in 
generic 
$D$ it is of the form $\frac{1}{n^{(D-3)/(D-2)}}$. Thus, if one verifies 
this for
$D=5$ for example, one can ascertain it will hold for arbitrary $D$.
We shall now briefly sketch our numerical procedure for finding
QN frequencies of a five dimensional black hole. This is just Leaver's and
Nollert's technique with minor modifications.

The QNMs of the higher-dimensional Schwarzschild black hole are characterized 
by the boundary conditions of incoming waves at the black hole horizon and 
outgoing waves at spatial infinity, written as
\begin{equation}
\Psi(r)\rightarrow\left\{
\begin{array}{ll}
e^{-i\omega r_*}&{\rm as}\ r_*\rightarrow \infty\\
e^{ i\omega r_*}&{\rm as}\ r_*\rightarrow-\infty\,, 
\end{array}
\right.
\label{def-bc}
\end{equation}
where the time dependence of perturbations has been assumed to be 
$e^{i\omega t}$.
%%%%%%%%%%%%%%%%%%%%%%%%%%%%%%%%%%%%%%%%%%%%%%%%%%%%%%%%%%%%%%%
\subsection{Perturbative calculation of QNMs of $D$-dimensional 
Schwarzschild black holes}
\label{apendice}
%%%%%%%%%%%%%%%%%%%%%%%%%%%%%%%%%%%%%%%%%%%%%%%%%%%%%%%%%%%%%
In this section we shall outline the procedure for computing the
first order corrections to the asymptotic value of the QN frequencies
of a $D$-dimensional Schwarzschild black hole.  This will be a 
generalization of Musiri and Siopsis' method \cite{musiri2}, so we
adopt all of their notation, and we refer the reader to their paper
for further details.
The computations are however rather tedious and the final expressions are too
cumbersome, so we shall refrain from giving explicit
expressions for the final result.

We start with the expansion of the potential $V$ near the singularity
$r=0$. One can easily show that near this point the potential 
(\ref{potentialj}) 
may be approximated as
\begin{equation}
V \sim -\frac{\omega ^2}{4z^2}\left[1-j^2+\frac{A}{(-z 
\omega)^{(D-3)/(D-2)}} \right]\,,
\label{potr0}
\end{equation}
where we adopted the conventions in \cite{musiri2} and therefore 
$z=\omega r_*$.
The constant $A$ is given by
\begin{eqnarray}
\!\!\!\!A\!&=&\frac{4l(l+D-3)+(D-2)(D-4)-(D-2)^2(1-j^2)}{(D-2)^{(D-1)/(D-2)}}+
\nonumber \\
& &
\frac{2(1-j^2)(D-2)^{(2D-5)/(D-2)}}{2D-5} \,.
\label{constA}
\end{eqnarray}
In $D=4$ this reduces to the usual expression
\cite{musiri2,andy} for the potential near $r=0$.  Expression
(\ref{potr0}) is a formal expansion in
$\frac{1}{\omega^{(D-3)/(D-2)}}$, so we may anticipate that indeed the
first order corrections will appear in the form
$\frac{1}{n^{(D-3)/(D-2)}}$.  So now we may proceed in a direct
manner:
we expand the wavefunction to first order in $\frac{1}{\omega^{(D-3)/(D-2)}}$ 
as
\begin{equation}
\Psi= \Psi^{(0)}+\frac{1}{\omega^{(D-3)/(D-2)}}\Psi^{(1)}\,,
\label{expansionpsi}
\end{equation}
and find that the first-order correction obeys the equation
\begin{equation}
\frac{d^2\Psi^{(1)}}{dz^2}+(\frac{1-j^2}{4z^2}+1)\Psi^{(1)}=
\omega^{(D-3)/(D-2)}\delta V \Psi^{(0)}\,,
\label{firstordereq}
\end{equation}
with 
\begin{equation}
\delta V=-\frac{A}{\omega^{(D-3)/(D-2)}(-z)^{(D-1)/(D-2)}}\,.
\label{deltaV}
\end{equation}
All of Musiri and Siopsis' expressions follow directly to the $D$-dimensional
case, if one makes the replacement $\sqrt{-\omega r_0} \rightarrow 
\omega^{(D-3)/(D-2)}$
in all their expressions. 
For example, the generalsolution of (\ref{firstordereq}) is
\begin{equation}
\Psi^{(1)}_{\pm}={\cal C} \Psi^{(0)}_{+} \int_{0}^z \Psi^{(0)}_{-}\delta V 
\Psi^{(0)}_{\pm}-
{\cal C} \Psi^{(0)}_{-} \int_{0}^z \Psi^{(0)}_{+}\delta V \Psi^{(0)}_{\pm}
\,,
\label{solfirstorder}
\end{equation}
where
 \begin{equation}
{\cal C}=\frac{\omega^{(D-3)/(D-2)}}{\sin{j\pi/2}}\,,
\label{calC}
\end{equation}
and the wavefunctions $\Psi^{(0)}_{\pm}$ are 
\begin{equation}
\Psi^{(0)}_{\pm}=\sqrt{\frac{\pi z}{2}}J_{\pm j/2}(z)\,.
\label{psi0}
\end{equation}
The only minor modification is their formula (30).
In $D$-dimensions, it follows that
\begin{equation}
\Psi^{(1)}_{\pm}=z^{1 \pm j/2 +k} G_{\pm}(z)\,,
\label{psi1}
\end{equation} 
where $k=\frac{D-4}{2(D-2)}$, and $G_{\pm}$ are even analytic functions of $z$.
So we have all the ingredients to construct the first order corrections in the 
$D$-dimensional Schwarzschild geometry. Unfortunately the final expressions 
turn out to be quite cumbersome, and we have not managed to simplify them.
We have worked with the symbolic manipulator {\it Mathematica}. It is 
possible to 
obtain simple expression for any particular $D$, but apparently not for a 
generic
$D$. For generic $D$ we write 
\begin{equation}
\frac{\omega}{T_H}=\ln{(1+2\cos{\pi j})}+i(2n+1)\pi+
\frac{{\rm Corr_D}}{\omega^{(D-3)/(D-2)}}
\label{xxx}
\end{equation}
So the leading term is 
$\frac{\omega}{T_H}=\ln{3}+i(2n+1)\pi$, for scalar, gravitational tensor and 
gravitational tensor perturbations (and the same holds for gravitational 
scalar).
However, for electromagnetic perturbations (vector or scalar) in $D=5$, 
$1+2\cos{\pi j}=1+2\cos{\frac{2\pi}{3}}=0$. So there seem to be no QN 
frequencies
as argued by Motl and Neitzke \cite{motl2}.
We obtained for $D=4$ the following result for the correction coefficient
in (\ref{xxx}),
\begin{equation}
{\rm Corr_4}=- \frac {(1-i)\pi ^{3/2}(-1+j^2-3l(l+1))\frac{\cos{j\pi}}{2}
\Gamma[1/4]} 
{ 3(1+2\cos{j\pi})\Gamma[3/4]\Gamma[3/4-j/2]\Gamma[3/4+j/2] }\,,
\label{corrd4}
\end{equation}
which reduces to Musiri and Siopsis' expression. For $D=5$
it is possible also to find a simple expression for the
corrections:
\begin{eqnarray}
\!\!\!{\rm Corr_5}\! &=\!\!\!&\frac {i(9j^2-(24+20l(l+2)))\pi^{3/2}} 
{ 120\times 3^{1/3}\Gamma[2/3]\Gamma[2/3-j/2]\Gamma[2/3+j/2] }
\nonumber \\ 
& &
\times \Gamma[1/6] 
\frac{ \frac{e^{ij\pi}}{\sin{((1/3+j/2)\pi)}}-
\frac{1}{\sin{((2/3+j/2)\pi)}}}{-1+e^{ij\pi}}\,.
\label{corrd5}
\end{eqnarray}
The limits $j \rightarrow 0,2$ are well defined and yield
\begin{equation}
{\rm Corr_5}^{j=0}= \frac {(1-i\sqrt{3})(6+10l+5l^2)\pi^{3/2}\Gamma[1/6]} 
{ 45\times 3^{1/3}\Gamma[2/3]^3} \,,
\label{corrd5j0}
\end{equation}
\begin{equation}
{\rm Corr_5}^{j=2}= \frac {(-1+i\sqrt{3})(-3+10l+5l^2)\pi^{3/2}\Gamma[1/6]} 
{ 45\times 3^{1/3}\Gamma[-1/3]\Gamma[2/3]\Gamma[5/3]} \,,
\label{corrd5j2}
\end{equation}
We list in Tables \ref{tab:corrj0}-\ref{tab:corrj2} some values of this 
five dimensional
correction for some values of the parameter $j$ and $l$ and compare
them with the results extracted numerically.  The agreement is quite
good.  The code for extracting the generic $D$-dimensional corrections
is available from the authors upon request.
It should not come as a surprise that for $j=2/3$ the correction term blows up:
indeed already the zeroth order term is not well defined.
%%%%%%%%%%%%%%%%%%%%%%%%%%%%%%%%%%%%%%%%%%%%%%%%%%%%%%%%%%%%%%
\subsection{Numerical procedure and results }
\label{numtechniqueresults}
%%%%%%%%%%%%%%%%%%%%%%%%%%%%%%%%%%%%%%%%%%%%%%%%%%%%%%%%%%%%%%

\subsubsection{Numerical procedure }
\label{numtechnique}

 In order to numerically obtain the QN frequencies, 
in the present investigation, we make use of Nollert's method 
\cite{nollert}, since asymptotic behaviors of QNMs in the 
limit of large imaginary frequencies are prime concern in the present study.

In our numerical study, we only consider the QNMs of the Schwarzschild black 
hole in the five dimensional spacetime, namely the $D=5$ case. 
As we have remarked, this should be representative. 
The tortoise 
coordinate is then reduced to 
\begin{equation}
r_*=x^{-1}+{1\over 2x_1}\ln(x-x_1)-{1\over 2x_1}\ln(x+x_1)\,, 
\end{equation}
where $x=r^{-1}$ and $x_1=r_h^{-1}$. Here, $r_h$ stands for the horizon 
radius of the black hole. The perturbation function $\Psi$ may be expanded 
around the horizon as 
\begin{equation}
\Psi=e^{-i\omega x^{-1}}(x-x_1)^\rho(x+x_1)^\rho\sum_{k=0}^\infty 
a_k\left({x-x_1\over-x_1}\right)^k\,, 
\label{expansion}
\end{equation}
where $\rho=i\omega/2x_1$ and $a_0$ is taken to be $a_0=1$. The 
expansion coefficients $a_k$ in equation (\ref{expansion}) are determined 
via the four-term recurrence relation (it's just a matter of substituting
expression (\ref{expansion}) in the wave equation (\ref{eveq})), given by
\begin{eqnarray}
&&\alpha_0a_1+\beta_0a_0=0\,, \nonumber \\
&&\alpha_1a_2+\beta_1a_1+\gamma_1a_0=0\,, \\
&&\alpha_ka_{k+1}+\beta_ka_k+\gamma_ka_{k-1}+\delta_ka_{k-2}=0\,,
\ k=2,3,\cdots, \nonumber
\end{eqnarray}
where 
\begin{eqnarray}
\alpha_k&=& 2(2\rho+k+1)(k+1)\,, \nonumber \\
\beta_k &=&-5(2\rho+k)(2\rho+k+1)-l(l+2)-{3\over 4} \nonumber \\
&&-{9\over 4}\,(1-j^2) \,, \nonumber \\
\gamma_k&=& 4(2\rho+k-1)(2\rho+k+1)+{9\over 2}\,(1-j^2)\,, \nonumber \\
\delta_k&=&-(2\rho+k-2)(2\rho+k+1)-{9\over 4}\,(1-j^2)\,. \nonumber
\end{eqnarray}
It is seen that since the asymptotic form of the perturbations as 
$r_*\rightarrow\infty$ is written in terms of the variable $x$ as
\begin{equation}
e^{-i\omega r_*}=e^{-i\omega x^{-1}}(x-x_1)^{-\rho}(x+x_1)^{\rho}\,,
\label{expwr} 
\end{equation}
the expanded perturbation function $\Psi$ defined by equation 
(\ref{expansion}) automatically satisfy the QNM boundary conditions 
(\ref{def-bc}) if the power series converges for $0\le x\le x_1$.
Making use of a Gaussian elimination \cite{Le90}, we can reduce the 
four-term recurrence relation to the three-term one, given by
\begin{eqnarray}
&&\alpha'_0a_1+\beta'_0a_0=0\,, \nonumber \\
&&\alpha'_ka_{k+1}+\beta'_ka_k+\gamma'_ka_{k-1}=0\,, 
\ k=1,2,\cdots, 
\end{eqnarray}
where $\alpha'_k$, $\beta'_k$, and $\gamma'_k$ are given in terms of 
$\alpha_k$, $\beta_k$, $\gamma_k$ and $\delta_k$ by 
\begin{eqnarray}
\alpha'_k=\alpha_k,\quad \beta'_k=\beta_k,\quad \gamma'_k=\gamma_k, 
\quad {\rm for\ } k=0,1, 
\end{eqnarray}
and
\begin{eqnarray}
\alpha'_k&=&\alpha_k, \nonumber \\
 \beta'_k&=&\beta_k-\alpha'_{k-1}\delta_k/\gamma'_{k-1}\,, \\
\gamma'_k&=&\gamma_k-\beta'_{k-1}\delta_k/\gamma'_{k-1}\,,\quad 
{\rm for\ }k\ge 2\,. \nonumber  
\end{eqnarray}
Now that we have the three-term recurrence relation for determining the 
expansion coefficients $a_k$, the 
convergence condition for the expansion (\ref{expansion}), namely the 
QNM conditions, can be written in terms of the continued fraction 
as \cite{Gu67,Le85}
\begin{eqnarray}
\beta'_0-{\alpha'_0\gamma'_1\over\beta'_1-}
{\alpha'_1\gamma'_2\over\beta'_2-}{\alpha'_2\gamma'_3\over\beta'_3-}
...\equiv 
\beta'_0-\frac{\alpha'_0\gamma'_1}{\beta'_1-
\frac{\alpha'_1\gamma'_2}{\beta'_2-\frac{\alpha'_2\gamma'_3}{\beta'_3-...}}}
=0 \,,
\label{a-eq}
\end{eqnarray}
where the first equality is a notational definition commonly 
used in the literature for infinite continued 
fractions.
Here we shall adopt such a convention.
In order to use Nollert's method, with which relatively high-order 
QNM with large imaginary frequencies can be obtained, we have to know 
the asymptotic behaviors of $a_{k+1}/a_k$ in the limit of 
$k\rightarrow\infty$. According to a similar consideration as that by 
Leaver \cite{Le90}, it is found that 
\begin{eqnarray}
{a_{k+1}\over a_k}=1\pm2\sqrt{\rho}\,k^{-1/2}+
\left(2\rho-{3\over 4}\right)k^{-1}+\cdots\,, 
\label{exp-R}
\end{eqnarray}
where the sign for the second term in the right-hand side is chosen so as 
to be 
\begin{eqnarray}
{\rm Re}(\pm2\sqrt{\rho}) < 0\,. 
\end{eqnarray}
In actual numerical computations, it is convenient to solve the $k$-th 
inversion of the continue fraction equation (\ref{a-eq}), given by 
\begin{eqnarray}
&&\beta'_k-{\alpha'_{k-1}\gamma'_k\over\beta'_{k-1}-}
{\alpha'_{k-2}\gamma'_{k-1}\over\beta'_{k-2}-}\cdots
{\alpha'_0\gamma'_1\over\beta'_0} \nonumber \\
&=& {\alpha'_{k}\gamma'_{k+1}\over\beta'_{k+1}-}
{\alpha'_{k+1}\gamma'_{k+2}\over\beta'_{k+2}-}\cdots\,. \quad 
\label{a-eq2}
\end{eqnarray}
The asymptotic form (\ref{exp-R}) plays an important role in Nollert's 
method when the infinite continued fraction in the right-hand side of
equation (\ref{a-eq2}) is evaluated \cite{nollert}.

%%%%%%%%%%%%%%%%%%%%%%%%%%%%%%%%%%%%%%%%%%%%%%%%%%%%%%%%%%%%%%%%%%%%%
\subsubsection{Numerical results}
\label{numericalresults}
%%%%%%%%%%%%%%%%%%%%%%%%%%%%%%%%%%%%%%%%%%%%%%%%%%%%%%%%%%%%%%%%%%%%%%%
Using the numerical technique described in section \ref{numtechnique}
we have extracted the 5000 lowest lying QN frequencies for the five
dimensional Schwarzschild black hole, in the case of scalar, gravitational 
tensor
and gravitational vector perturbations.
For electromagnetic vector perturbations the situation is different:
the real part rapidly approaches zero, and we have not been able to compute
more than the first $30$ modes.

\bigskip\medskip
\centerline {\it (i) Low-lying modes}
\medskip

Low lying modes are important since they govern the intermediate-time
evolution of any black hole perturbation. As such they may play a role
in TeV-scale gravity scenarios \cite{hamed}, and higher dimensional
black hole formation \cite{vitoroscarjose,bhprod}.  For example, it is
known \cite{vitoroscarjose,cardosorad,bertihighd} that if one forms black holes
through the high energy collision of particles, then the fundamental
quasinormal frequencies serve effectively as a cuttof in the energy
spectra of the gravitational energy radiated away.
In Tables \ref{tab:qnf5dj0}-\ref{tab:qnf5dj23} we list the five lowest lying
QN frequencies for some values of the multipole index $l$.
The fundamental modes are in excellent agreement with the ones presented
by Konoplya \cite{konoplyawkb,roman2} using a high-order WKB approach.
Notice he uses a different convention so one has to be careful when 
comparing the results. For example, in our units Konoplya obtains for
scalar and tensorial perturbations ($j=0$) with $l=2$ a fundamental
QN frequency $\omega _{0}/T_H=9.49089+2.24721i$, whereas we get, from 
Table \ref{tab:qnf5dj0} the number $9.4914+2.2462i$ so the WKB approach
does indeed yield good results, at least for low-lying modes, since it
is known it fails for high-order ones.

\begin{table}
\caption{\label{tab:qnf5dj0} The first lowest QN frequencies
for scalar and gravitational tensor ($j=0$) perturbations 
of the five dimensional Schwarzschild black hole. The frequencies
are normalized in units of black hole temperature, so the Table
really shows $\frac{\omega}{T_H}$, 
where $T_H$ is the Hawking temperature of the black hole. }
\begin{ruledtabular}
\begin{tabular}{l|lll}  \hline
\multicolumn{1}{c}{} &
\multicolumn{3}{c}{ $j=0$}\\ \hline
$n$ &$l=0$:         &$l=1$:      &$l=2$\\ \hline
0&3.3539+2.4089i &6.3837+2.2764i &9.4914+2.2462i  \\ \hline 
1&2.3367+8.3101i &5.3809+7.2734i &8.7506+6.9404i  \\ \hline 
2&1.8868+14.786i &4.1683+13.252i &7.5009+12.225i \\ \hline 
3&1.6927+21.219i &3.4011+19.708i &6.2479+18.214i \\ \hline 
4&1.5839+27.607i &2.9544+26.215i &5.3149+24.597i \\ \hline 
\end{tabular}
\end{ruledtabular}
\end{table}

\begin{table}
\caption{\label{tab:qnf5dj2} The first lowest QN frequencies
for gravitational vector ($j=2$) perturbations 
of the five dimensional Schwarzschild black hole. The frequencies
are normalized in units of black hole temperature, so the Table
really shows $\frac{\omega}{T_H}$, 
where $T_H$ is the Hawking temperature of the black hole. }
\begin{ruledtabular}
\begin{tabular}{l|lll}  \hline
\multicolumn{1}{c}{} &
\multicolumn{3}{c}{ $j=2$}\\ \hline
$n$ &$l=2$:       &$l=3$:           &$l=4$\\ \hline
0&7.1251+2.0579i  &10.8408+2.0976i  &14.3287+2.1364i  \\ \hline 
1&5.9528+6.4217i  &8.8819+1.0929i   &12.8506+1.0574i   \\ \hline 
2&3.4113+12.0929i &7.2160+16.3979i  &11.5069+16.0274i   \\ \hline 
3&2.7375+19.6094i &5.5278+22.3321i  &9.99894+21.5009i  \\ \hline 
4&2.5106+26.2625i &3.9426+28.6569i  &8.56386+27.4339i  \\ \hline 
\end{tabular}
\end{ruledtabular}
\end{table}
\vskip 1mm

\vskip 1mm
\begin{table}
\caption{\label{tab:qnf5dj23} The first lowest QN frequencies
for electromagnetic vector ($j=2/3$) perturbations 
of the five dimensional Schwarzschild black hole. The frequencies
are normalized in units of black hole temperature, so the Table
really shows $\frac{\omega}{T_H}$, 
where $T_H$ is the Hawking temperature of the black hole. }
\begin{ruledtabular}
\begin{tabular}{l|lll}  \hline
\multicolumn{1}{c}{} &
\multicolumn{3}{c}{ $j=2/3$}\\ \hline
$n$ &$l=1$:        &$l=2$:           &$l=3$\\ \hline
0&5.9862+2.2038i &9.2271+2.2144i &12.4184+2.2177i\\ \hline 
1&4.9360+7.0676i &8.4728+6.8459i &11.8412+6.7660i \\ \hline 
2&3.6588+12.9581i&7.1966+12.0804i&10.7694+11.6626i\\ \hline 
3&2.8362+19.3422i&5.9190+18.0343i&9.4316+17.1019i \\ \hline 
4&2.3322+25.7861i&4.9682+24.3832i&8.1521+23.0769i  \\ \hline 
\end{tabular}
\end{ruledtabular}
\end{table}

In the four dimensional case, and for gravitational vector 
perturbations, there is for is each $l$, a purely
imaginary QN frequency, which had been coined an ``algebraically
special frequency'' by Chandrasekhar \cite{chandra}.  For further
properties of this special frequencies we refer the reader to
\cite{brinkspecial}. The existence of these purely imaginary
frequencies translates, in the four dimensional case, a relation
between the two gravitational wavefunctions (i.e., between the
Regge-Wheeler and the Zerilli wavefunction, or between the
gravitational vector and gravitational scalar wavefunctions,
respectively).  It is
possible to show for example that the associated potentials are
related through supersymmetry.  Among other consequences, this
relation allows one to prove that the QN frequencies of both
potentials are exactly the same \cite{chandra}.

We have not spotted any purely imaginary QN frequency for this
five-dimensional black hole.  In fact the QN frequency with the
lowest real part is a gravitational vector QN frequency with $l=4$ and
overtone number $n=13$, $\omega=8.7560\times 10^{-2}+6.9934i$.  This
may indicate that in five dimensions, there is no relation between the
wavefunctions, or put another way, that the potentials are no longer
superpartner potentials.  This was in fact already observed by Kodama
and Ishibashi \cite{kodama} for any dimension greater than four. It
translates also in Tables \ref{tab:qnf5dj0}-\ref{tab:qnf5dj2},
which yield different values for the QN frequencies. Although we have not
worked out the gravitational scalar QN frequencies, a WKB approach
can be used for the low-lying QN frequencies, and also yields different values
\cite{konoplyawkb,roman2}.

\bigskip\medskip
\centerline {\it (ii) Highly damped modes}
\medskip

\begin{figure}
\centerline{\includegraphics[width=7 cm,height=7 cm]
{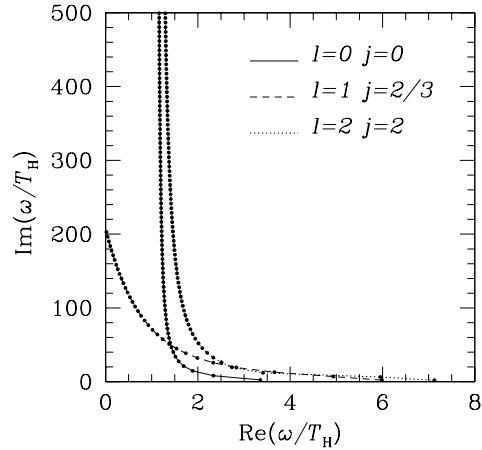}}
\caption{The QN frequencies of scalar ($j=0$), gravitational vector ($j=2$) 
and electromagnetic vector ($j=2/3$) perturbations
of a five-dimensional Schwarzschild black hole. We only show the 
lowest radiatable multipoles,
but the asymtptoic behaviour is $l$-independent.}
\label{fig:qnall5d}
\end{figure}

\begin{figure}
\centerline{\includegraphics[width=7 cm,height=7 cm]
{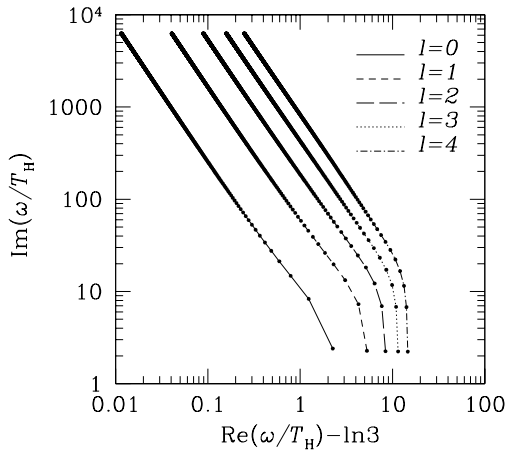}}
\caption{The QN frequencies of 
scalar and gravitational tensor perturbations ($j=0$) 
of a five-dimensional Schwarzschild black hole.
Notice that for frequencies with a large imaginary part
the slope is $l$-independent and equals $3/2$.}
\label{fig:qntensor5d}
\end{figure}

\begin{figure}
\centerline{\includegraphics[width=7 cm,height=7 cm]
{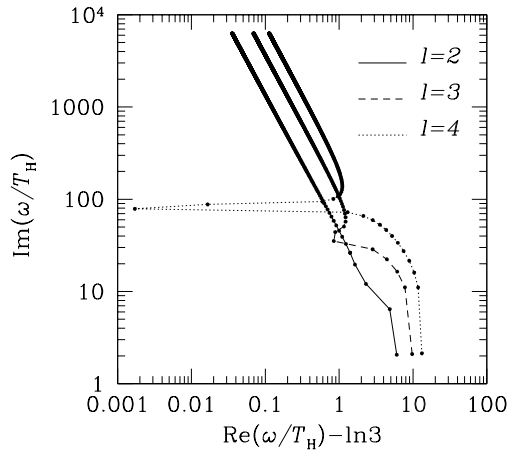}}
\caption{The QN frequencies of gravitational vector perturbations
($j=2$) of a five-dimensional Schwarzschild black hole.}
\label{fig:qnvector5d}
\end{figure}

Our numerical results for the highly damped modes, i.e., QN
frequencies with a very large imaginary part, are summarized in
Figures \ref{fig:qnall5d}-\ref{fig:qnvector5d} and in Tables
\ref{tab:corrj0}-\ref{tab:corrj2}.

\vskip 1mm

In Fig. \ref{fig:qnall5d} we show our results for scalar,
gravitational tensor ($j=0$), gravitational vector ($j=2$) and
electromagnetic vector ($j=2/3$) QN frequencies.  For $j=0\,,2$ the
real part approaches $\ln{3}$ in units of the black hole temperature, a
result consistent with the analytical result in \cite{motl2} (see also
section \ref{apendice}).  Electromagnetic vector ($j=2/3$) QN
frequencies behave differently: the real part rapidly approaches zero,
and this makes it very difficult to compute the higher modes.
In fact, the same kind of problem appears in the Kerr geometry \cite{qnmkerr},
and this a major obstacle to a definitive numerical characterization of the
highly damped QNMs in this geometry. 
We have
only been able to compute the first $30$ modes with accuracy.  The
prediction of \cite{motl2} for this case (see section \ref{apendice})
is that there are no asymptotic QN frequencies.
It is left as an open question whether this is true or not.
Our results indicate that the real part rapidly approaches zero, but we
cannot say whether the modes die there or not, or even if they perform
some kind of oscillation. The imaginary part behaves in the usal manner,
growing linearly with mode number.

Let us now take a more detailed look at our numerical data for
scalar and gravitational QN frequencies.
In Fig. \ref{fig:qntensor5d} we show in a $\ln$ plot
our results for scalar and gravitational tensor ($j=0$) QN frequencies.
\vskip 1mm
\begin{table}
\caption{\label{tab:corrj0} The correction coefficients for the
five-dimensional Schwarzschild black hole, both numerical, here
labeled as ``${\rm Corr_{5}^{N}}$'' and analytical, labeled as ``${\rm
Corr_{5}^{A}}$''. These results refer to scalar or gravitational tensor
perturbations ($j=0$).
The analytical results are extracted from the
analytical formula (\ref{corrd5j0}).  Notice the very good agreement, to
within 0.5\% or less, between the numerically extracted results and the
analytical prediction.}
\begin{ruledtabular}
\begin{tabular}{l|ll}  \hline
\multicolumn{1}{c}{} &
\multicolumn{2}{c}{ $j=0$}\\ \hline
$l$ &${\rm Corr_5^N}$:&${\rm Corr_5^A}$\\ \hline
0   &  1.155 - 1.993i &  1.15404-1.99886i \\ \hline 
1   &  4.058 - 6.956i &  4.03915-6.99601i \\ \hline 
2   &  8.894 - 15.26i &  8.84765-15.3246i \\ \hline 
3   &  15.64 - 26.79i &  15.5796-26.9847i \\ \hline 
4   &  24.33 - 41.67i &  24.2349-41.9761i \\ \hline 
\end{tabular}
\end{ruledtabular}
\end{table}
\vskip 1mm

Our numerical results are very clear: asymptotically the QN
frequencies for scalar and gravitational tensor perturbations $(j=0)$ behave
as
\begin{equation}
\frac{\omega}{T_H}=\ln{3}+i\,(2n+1)\pi+\frac{{\rm Corr_5}}{n^{2/3}}\,\,,
\label{asympt5dScalar}
\end{equation}
So the leading terms $\ln{3}+(2n+1)i$ are indeed the ones predicted
in \cite{motl2}. Interestingly, the first corrections do not appear as
$\frac{1}{\sqrt{n}}$, but as $\frac{1}{n^{2/3}}$. This was shown to be
the expected analytical result in section \ref{apendice}, where we
generalized Musiri and Siopsis' \cite{musiri2} results to higher dimensions.
In table
\ref{tab:corrj0} we show the coefficient ${\rm Corr_5}$ extracted
numerically along with the predicted coefficient (see expression
(\ref{corrd5j0})).  The table is very clear: the numerical values
match the analytical ones.  Another confirmation that the corrections
appear as $\frac{1}{n^{2/3}}$ is provided by
Fig. \ref{fig:qntensor5d}. In this figure the QN frequencies are plotted
in a $\ln$ plot for an easier interpretation. One sees that for large
imaginary parts of the QN frequencies, the slope of the plot is
approximately $-3/2$, as it should be if the corrections are of the
order $\frac{1}{n^{2/3}}$.

In Fig. \ref{fig:qnvector5d} we show in a $\ln$ plot our results for
gravitational vector ($j=2$) QN frequencies.  Again, vector QN
frequencies have the asymptotic behaviour given by expression
(\ref{asympt5dScalar}), with a different correction term ${\rm
Corr_5}$. Again, the $\frac{1}{n^{2/3}}$ corrections show themselves
in the $\ln$ plot of Fig. \ref{fig:qnvector5d}: for very large imaginary
parts, the slope is $-3/2$, as it should.  The numerically extracted
coefficient ${\rm Corr_5}$ for $j=2$ perturbations is listed in Table
\ref{tab:corrj2}, along with the analytically predicted value (see
section \ref{apendice}).  the agreement is very good.
\vskip 1mm
\begin{table}
\caption{\label{tab:corrj2} The correction coefficients for the
five-dimensional Schwarzschild black hole, both numerical, here
labeled as ``${\rm Corr_{5}^{N}}$'' and analytical, labeled as ``${\rm
Corr_{5}^{A}}$'', for gravitational vector perturbations ($j=2$).  
The analytical results are extracted from the
analytical formula (\ref{corrd5j2}).  Notice the very good agreement, to
within 0.5\% or less, between the numerically extracted results and the
analytical prediction.}
\begin{ruledtabular}
\begin{tabular}{l|ll}  \hline
\multicolumn{1}{c}{} &
\multicolumn{2}{c}{ $j=2$}\\ \hline
$l$ &${\rm Corr_5^N}$:&${\rm Corr_5^A}$:\\ \hline
2   & 3.568 - 6.139i  & 3.5583 -6.16316i \\ \hline 
3   & 6.947 - 11.94i  & 6.92425-11.9932i \\ \hline 
4   & 11.29 - 19.38 i & 11.2519-19.4889i \\ \hline 
\end{tabular}
\end{ruledtabular}
\end{table}
\vskip 1mm

%%%%%%%%%%%%%%%%%%%%%%%%%%%%%%%%%%%%%%%%%%%%%%%%%%%%%%%%%%%%%%%%%%%%
\section{Discussion of results and future directions}
%%%%%%%%%%%%%%%%%%%%%%%%%%%%%%%%%%%%%%%%%%%%%%%%%%%%%%%%%%%%%%%%%%%%%
We have made an extensive survey of the QNMs of the four and five
dimensional Schwarzschild black hole.  The investigation presented
here makes a more complete characterization of the highly damped QNMs in the
Schwarzschild geometry.  In the four-dimensional case, we confirmed
previous numerical results regarding the scalar and gravitational
asymptotic QN frequencies. We found that both the leading behaviour
and the first order corrections for the scalar and gravitational
perturbations agree extremely well with existing analytical formulas.  We
have presented new numerical results concerning electromagnetic QN frequencies
of the four-dimensional Schwarzschild geometry. In particular, this is
the first work dealing with highly damped electromagnetic QNMs.
Again, we find that the leading behaviour and the first order
corrections agree with the analytical calculations. The first order
corrections appear at the $\frac{1}{n^{3/2}}$ level.
In the five-dimensional case this represents the first study on highly
damped QNMs.  We have seen numerically that the asymptotic behaviour
is very well described by $\frac{\omega}{T_H} = \ln{3}+\pi i(2n+1)$ for
scalar and gravitational perturbations, and agrees with the predicted
formula.  Moreover the first order corrections appear at the level
$\frac{1}{n^{2/3}}$, which is also very well described by the
analytical calculations, providing one more consistency check on the
theoretical framework.  
In generic $D$ dimensions the corrections appear at the
$\frac{1}{n^{(D-3)/(D-2)}}$ level.
We have not been able to prove or disprove the analytic result
for electromagnetic QN frequencies in five dimensions ($j=2/3$).
Other conclusions that can be taken from
our work are: the monodromy method by Motl and Neitzke is correct.
It is important to test it numerically, as we did, since there are some 
ambiguous 
assumptions in that method. Moreover, their method is highly flexible, 
since it allows an
easy computation of the correction terms, as shown by Musiri and
Siopsis, and generalized here for the higher dimensional Schwarzschild
black hole.
We have basically showed numerically that the monodromy method
by Motl and Neitzke \cite{motl2}, and its extension by Musiri and
Siopsis \cite{musiri2} to include for correction terms in overtone
number, are excellent techniques to investigate the highly damped QNMs
of black holes. 

Highly damped QNMs of black holes are the bed rocks that the recent 
conjectures
\cite{hod,dreyer,ling} are built on, relating these to black hole area
quantization.  However, to put on solid ground the conjecture that
QNMs can actually be of any use to black hole area quantization, one
has to do much better then one has been able to do up to now.  In
particular, it is crucial to have a deeper understanding of the 
Kerr and Reissner-Nordstr\"om QNMs. 
These are, next to the Schwarzschild, the simpler asymptotically
flat geometries. If these conjectures hold, they must to do also
for these spacetimes.
Despite the fact that there are convincing numerical results for the Kerr 
geometry
\cite{qnmkerr}, there seems to be for the moment no serious analytical
investigation of the highly damped QNMs for this spacetime.  Moreover,
even though we are in possession of an analytical formula for the
highly damped QNMs of the Reissner-Nordstr\"om geometry \cite{motl2},
and this has been numerically tested already \cite{bertirn}, we have
no idea what it means!  In particular, can one use it to quantize the
black hole area? In this case the electromagnetic perturbations
are coupled to the gravitational ones. How do we proceed with the
analysis, that was so simple in the Schwarzschild geometry? 
Is the assumption of equally spaced eigenvalues a correct one, in this case?
Can these ideas on black hole area quantization be translated to 
non-asymptotically 
flat spacetimes, as the
de Sitter or anti-de Sitter spacetime?
These are the fundamental issues that remain to be solved in this
field. Should any of them be solved satisfactorily, then these
conjectures would gain a whole new meaning. As they stand, it may just
be a numerical coincidence that the real part of the QN frequency goes
to $\ln{3}$.

%%%%%%%%%%%%%%%%%%%%%%%%%%%%%%%%%%%%%%%%%%%%%%%%%%%%%%%%%%%%
\section*{Acknowledgements}
The authors would like to thank Emanuele Berti and Alejandro Corichi for very useful
comments and suggestions. The authors thank also Ricardo Schiappa and Jos\'e Nat\'ario
for very useful discussions, and are specially indebted to Roman
Konoplya for sharing his numerical results prior to publication.
This work was partially funded by Funda\c c\~ao para a Ci\^encia e
Tecnologia (FCT) - Portugal through project PESO/PRO/2000/4014 
and CERN/FIS/43797/2001. SY
acknowledges finantial support from FCT through project SAPIENS
36280/99.  VC also acknowledge finantial support from FCT
through PRAXIS XXI programme.  JPSL 
thanks Observat\'orio Nacional do Rio de Janeiro
for hospitality. 

%%%%%%%%%%%%%%%%%%%%%%%%%%%%%%%%%%%%%%%%%%%%%%%%%%%%%%%%%%%

\end{document}